# Peaked structures in noise power spectra as signature of avalanche correlation.

E. Celasco (1),  M. Celasco (1),  R. Eggenhöffner (1)

( (1) Physics Department, Università di Genova, via Dodecaneso 33, Genova, Italy,
(2) Physics Department, Politecnico di Torino, C.so Duca degli Abruzzi 24, Torino, Italy)

ABSTRACT

An outstanding topic on noise phenomena is the occurrence of peaked structures in many natural systems in a wide range $10^{-1} \div 10^{6}$ Hz. All existing theories failed to explain this issue. The present theory based on first principle statistics of elementary events clustered in time-amplitude correlated large avalanches leads to a noise spectral power master equation suitable for any peaked noise spectra. The excellent agreement with our current noise experiments in high $T_c$ superconductors in the dendritic regime and with optical noise experiments in E.coli demonstrates firstly that avalanche correlation is the physical source of spectral peaks.

*INTRODUCTION*

The challenging interest on noise power spectra has recently oriented to the spectral peak features superimposed to the *1/f* behavior. Experiments showing these noise peaked behavior concern a large variety of physical systems i.e. superconductors at very low temperatures ($T<<T_c$) in the dendritic regime [1,2], single electron tunneling oscillations detected by SET [3], archetypal biological systems (notably a single cell bacteria [4,5]), polymer motion in shear flow [6], slide friction in metals [7], raw speech signals [8] and many other systems showing crackling noise of the Barkhausen (B) type observed in ferromagnets [9]. Peaked structures have a quite different origin with respect to the change of slopes in the *1/f* behavior which have been studied for a long and associated generally to clustering of elementary events [10] or to creation-annihilation events as, for instance, in vacancy fluctuations in metals [11]. However, until now, a comprehensive interpretation of the observed peak deviations from the usual *1/f* behavior in the framework of universal power law models was attempted unsuccessfully. In the present work we are focusing our analyses on the role of the correlation among large clustering (avalanches or bundles) as the main source of the noise peaks experimentally observed.

The appearance of a distinct peaked structure was first reported [12] in a tube of $Nb_{0.47}Ti_{0.53}$ superconductor at $T<<T_c$ around 200 Hz in driving magnetic field sweeps. Authors stressed that in this strongly driven system the critical behavior cannot be achieved and that the self-organized criticality (SOC) theory cannot be applied. In high $T_c$ superconductors, noise peaks at 15-180 Hz were observed at 4.2K in both $MgB_2$ films [2] and YBCO bulks [13] in the same experimental conditions promoting the thermo-magnetic instabilities with dendritic structures observed in magneto-optical and Hall effect experiments [1]. A similarly intriguing effect is observed in the noise spectra of single cell Escherichia coli (E.coli) bacteria: a peak from the dynamics of clock and anticlockwise spinning rotations of flagellar motors activated by a protein is obtained below 1 Hz [4,5]. An adopted two-state model [5] failed to explain these peaked noise spectra since it neglected the correlation among the duration times in the bacteria motors. Evidence of a peak at about 60 Hz is claimed in the frictional fluctuations of a sliding aluminum load on a $1/f^\gamma$ behavior with $\gamma=1.19 \div 1.37$ depending upon the load [7]. A SOC model invoked also here doesn't account for the peak feature. Single electron tunneling in a several junctions array was shown to give spectral peaks at frequencies above 1 MHz, originating from the current and temperature broadening of the $f = I/e$ signal. The response of human neurons to auditorial and visual signals exhibits both slope changes and peaks in the frequency range up to about 10 kHz.

The deviations from the power-law noise dependence in ferromagnetic materials under magnetic field sweeps, i.e. the Barkhausen noise (BN) have been studied for a long [14-18], with recent advancements [19,20]. When time-amplitude correlation among large discontinuities (avalanches) is involved, a drop was obtained in the low frequency range [16]. The noise peak dominates the $1/f^\gamma$ spectrum in the BN whereas peak and $1/f^\gamma$ appears often of comparable intensity in the systems of recent interest.

The experimental spectra with a sharp peak appear composed by two contributions, a $1/f^\gamma$ and a peak, with different relative amplitudes. In the present approach, we assume that noise is given by sequences of elementary events, by elementary events correlated in a large avalanche and by correlation among avalanches. The $1/f^\gamma$ can account for the first two noise sources and the peak for the latter.

## THE STATISTICAL MODEL

The burst signals from many natural systems can be represented as sequences of non-independent overlapping pulses having random shape and amplitude. The basic idea is to calculate the power spectrum of such signals in the framework of a statistical model. Following the pioneering works by P.Mazzetti [14] and C.Heiden [15], pulses with suitable correlation between the events give power spectrum explaining quantitatively ferromagnetic domain dynamics in the context of the BN effect. Correlation between the pulse parameters (duration, amplitude and period between successive pulses) was employed to explain the complex dynamical behavior of different physical systems such as the clustering of elementary events. Independently on the specific physical origin of the phenomena studied, a distribution function of elementary events can be attempted. BN was explained [21] by assuming a simple Poissonian distribution for independent events in each cluster and to introduce the correlation among elementary events in the analytical expression of the power spectrum. The main result is summarized in the following equation of the power spectrum $\Phi(\omega)$:

$$1) \quad F(\omega) = \frac{2r j(\omega)}{1 + \omega^2 r^2 t_0^2}$$

in which $j(\omega)$ (flat in the frequency range of experimental interest) is the power spectral function of completely independent events, $r$ is the average number of pulses in each avalanche, $t_0$ the average time period between two subsequent events.

Further, a time-amplitude correlation between subsequent large B discontinuities was introduced to account for the drop in the noise at low frequencies given by large clustered domain transitions [16]. The elementary B pulses were assumed to cluster into large discontinuities and their number to be proportional to the time interval between two large subsequent B discontinuities. A linear dependence was assumed between the average pulse number $n(x)$ clustered in each avalanche of frequency $\nu_g$ (Poisson distributed $P(x) = n_g e^{-n_g x}$ with the $x$–time interval between two large avalanche discontinuities). Within these assumptions, the power spectrum reads [16]:

$$2) \quad \Phi(\omega) = n_g \left\{ <\overline{|S(\omega,x)|^2}> + 2\mathrm{Re}\left( \overline{<|S(\omega,x)|>} \frac{\int \overline{S(\omega,x)} P(x) e^{j\omega x} dx}{1 - \int P(x) e^{j\omega x} dx} \right) \right\}$$

where $\overline{S(\omega,x)}$ is the Fourier transform of an average B large discontinuity and $<\overline{S(\omega,x)}>$ its average over the x-time distribution adopted. A large discontinuity can be modeled by $S(\omega,x) = a n(x) e^{j\psi_0}$ being $a$ an average avalanche spectral area and $\psi_0$ a constant phase factor. In the BN case, the power spectrum becomes:

$$3) \quad \Phi(\omega) = \frac{2r j(\omega)}{1 + \omega^2 r^2 t_0^2} \left\{ 1 - \frac{1 + \omega^2 r^2 t_0^2}{(\omega^2 + n_g^2)} n_g^2 \right\}$$

With respect to eqn.1, the factor in curl parenthesis gives the $\omega^2$-drop at frequencies $\omega \ll \nu_g$. At large frequencies the leading term has the $\omega^{-2}$ behavior as eqn.1. A peak results from the two combined behaviors. The low frequency drop in the power spectrum calculations as invoked above in our statistical model can be achieved with different approaches. Heiden [15] also obtained a $\omega^2$ dependence in the low frequency range assuming the correlation between the parameters of elementary events whereas in eqn.2 correlation is established among avalanches. Later, starting from a dynamical analysis of the Bloch walls in ferromagnets, a spectral dependence identical to eqn.3 in the limit $r^2 t_0^2 n_g^2 \ll 1$ was obtained [17], a limit suitable for B processes in which time-amplitude correlation is essentially of magnetostatic origin. A fully comprehensive treatment of noise processes associated with avalanche dynamics regime is required: in superconductors, for instance, thermomagnetic instabilities of dendritic structures at the origin of avalanches are affected also by transport current.

In the present work, to explain the peaked structures reported by recent experiments, we introduce a scaling relation between average pulse number $n(x)$ (i.e. average size avalanche) and the time duration following the generalized behavior $n(x) \propto x^s$. From eqn.2, the generalized global power spectrum equation reads:

$$4) \quad \Phi(w) = \frac{2 r j(w)}{1 + w^2 r^2 t_0^2} \left\{ 1 - \frac{1 + w^2 r^2 t_0^2}{(w^2 + n_g^2)} n_g^2 c_s(w) \right\}$$

In eqn.4, the time-amplitude correlation term is modulated by the function $\chi_s(\omega)$ reported in Fig.1:

$$5) \quad c_s(w) = n_g^{s-1} \frac{\sin(s \, \text{arctg} \, w/n_g)}{w(w^2 + n_g^2)^{s/2 - 1}} f(\Gamma)$$

Complex integrals of the type $\int_0^\infty z^s e^{-z} dz$ and $\int_0^\infty z^{2s} e^{-z} dz$ with $z = (n_g - jw)$ enter the analytical development to obtain the generalized eqns.4-5, as well as the factor $f(\Gamma) = 2\Gamma^2(s+1)/\Gamma(2s+1)$ of the Gamma functions $\Gamma$.

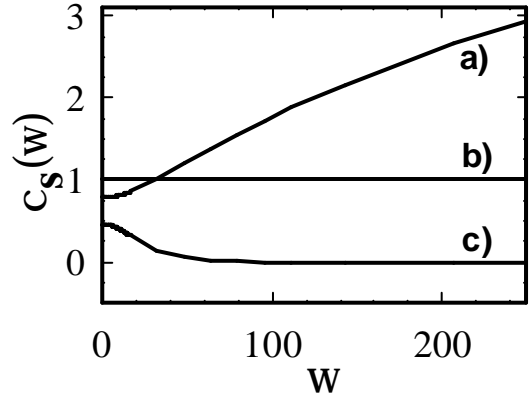

FIG.1 – Behaviour of the modulation function $\chi_s(\omega)$ for three values of the parameter
*s:* a) 0.5, b) 1, c) 2.5 at $\nu_g$=200 Hz vs. the angular velocity $\omega$(rad/s).

When avalanche correlation is disregarded (*s=0*), $c_s$ is identically zero at every frequency and eqn.1, consistent with this condition, is obtained. In the case of linear correlation between avalanches (*s=1*), $c_s=1$: in these conditions eqn.3 is recovered as expected. More generally, as shown in Fig.1, *0<s<1* turns out suitable for systems in which correlation increases with frequency whereas *s>1* for correlation decrease at increasing frequency.

A second result from our generalization is that we provide a frequency dependence around the peak more adaptable to the large variety of physical systems. Also the different correlation mechanisms related to different ordering state of the material (crystalline or amorphous) can be taken into account by the appropriate choice of the *s* parameter. Third point, it explains the deviations from the $1/f^g$ behavior with few disposable parameters, i.e. $\rho\varphi_0$ (the limit of $\varphi(\omega)$ when $\omega\rightarrow 0$), $\rho\tau_0$, $\nu_g$ and the power parameter *s*. However, $\nu_g$ and $\rho\varphi_0$ are related to peak frequency and amplitude, respectively: only s and $\rho\tau_0$ are the disposable parameters of this treatment. Thus, starting from first principles ruling the superposition of elementary events, a generalized description of peaks in noise spectra unconstrained to the microscopical origin of the events is obtained.

## *RESULTS ON HIGH TC SUPERCONDUCTORS IN THE DENDRITIC REGIME*

In high $T_c$ superconductors, peaked structures superimposed on the $1/f$ noise behavior were firstly detected in our experiments with a low $T_c$ DC-SQUID both in thin films and in bulks at $T<<T_c$ [2,13]. The samples are in weak thermal contact with the liquid helium reservoir in order to allow the thermomagnetic instabilities to play the dominant role in the onset of a dendritic regime. The power spectral density measured in a thin film of "metallic" $MgB_2$ and in a ceramic YBCO bulk are reported in Fig.2 (a) and (b), respectively where the peak structures are specifically evidenced.

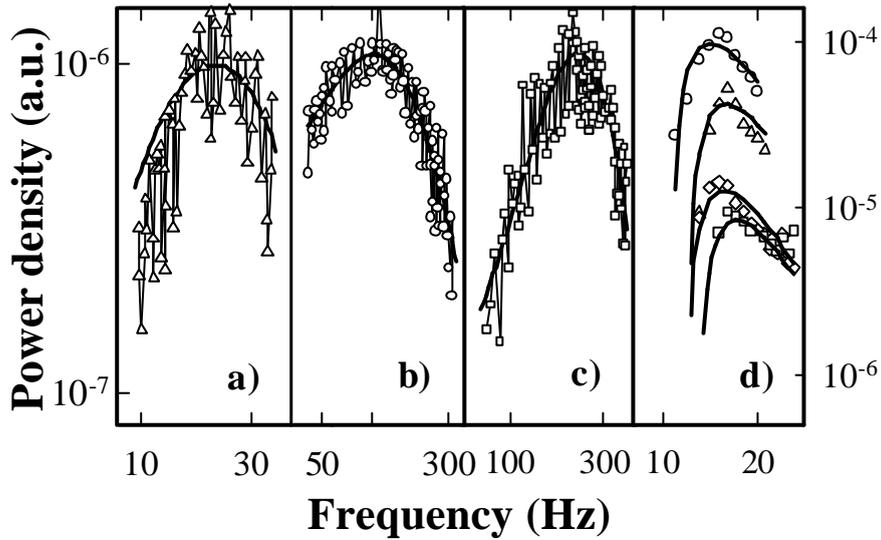

FIG.2 – Frequency dependence of power density at 4.2 K in superconducting $MgB_2$ thin film **(a), (b), (c)** (left scale) and YBCO bulk sample **(d)** (right scale). Experimental peaks (curves with symbols) are obtained subtracting from the measured power spectra the $1/f^g$ behaviour at frequencies below the peak, the white noise at higher frequencies and a polynomial connection between the two regimes at frequencies around the peak. Combined transport current densities j and magnetic field values B are adopted to enhance noise peaks by correlation among avalanches promoted by thermal instabilities. In Fig 2 (a) $\Delta$: B=200 mT, j=5.0 $GA/m^2$; in (b) O: B=450 mT, j=2.5 $GA/m^2$, the combination with highest peak in $MgB_2$; in (c) : B=450 mT, j=1.0 $GA/m^2$. The curves (a) $\div$ (c) are scaled to the same peak amplitude. In Fig.2 (d), at the fixed current of j=8 $kA/m^2$, $\Delta$: B=140 mT, O: B=250 mT, the highest peak in YBCO; : B=400 mT; $\Diamond$: B=600 mT. The curves O $\div$ in Fig.2 (d) are staked for clarity. Thick solid curves are obtained from equation (2): fit parameters in the ranges s $\approx$ 0.5, $\rho\tau_0$ = 0.8 $\div$ 2 ms ($MgB_2$) and s = 2.5 $\div$ 2.8 and $\rho\tau_0$= 100 $\div$ 500 ms (YBCO) are obtained.

The low frequency behavior reported previously [2,13] of the $1/f^g$ gives a **g** parameter close to the limiting value **g**=2, a stringent evidence of clustering of elementary pulses [10]. A first attempt to fit the data reported in Fig.2 (a), (b) was performed adopting with *s=1* in eqn.4,5 as suggested by the treatment on the BN effect in textured polycrystalline Fe-Si magnets [16]. A qualitative agreement is obtained even if the fine structure of the peaks is not entirely reproduced within this assumption. Also the quadratic scaling ($s=2$) suggested in connection with amorphous Fe-Co-B magnetic alloys [18] leads to even more broadened peaks with respect to our experimental data. With respect to the B effect, physics is quite different in the present superconducting systems in the dendritic regime at very low temperatures. The dendritic formation from thermomagnetic instabilities give rise to avalanche dynamics [1,2]: in $MgB_2$ thin films, the dynamics at small time scales appear relevant in view of the fast vortex-antivortex interactions and annihilations. In this frame, a time-amplitude correlation with $0<s<1$ seems more suitable and the value $s=0.5$ is selected in eqn.4 as a best-fit value, as shown by the solid curves reported in Fig.2 (a). A satisfactory agreement is observed, in particular both the peak amplitude and position as well as the frequency dispersion is correctly accounted for by best-fit parameters in the generalized equation. In bulk ceramic YBCO, noise was investigated at feeding current close to the critical value and in different magnetic fields up to 600 mT. As shown in Fig.2 (b), the best attempts to reproduce the experimental peak structure achieved satisfactory agreement with the choice $s=2.5$¸$2.8$, pointing out the relevance of slower dynamical processes in a bulk ceramic superconductor with respect to $MgB_2$ film.

### *RESULTS ON E.COLI SINGLE CELL SPECTRA.*

Among the experimental noise spectra appeared recently on complex systems of biophysical [4,5] or biochemical interest [6] showing pronounced peaked structures, we focus in the present work on the well known bacterium E.coli. Although a great deal of works has been devoted to study biophysical properties, only recently the dynamical behavior of single cells was deeply investigated. Concerning the translation and rotation dynamics, the cell is propelled by several flagella activated by motors each one having two possible clockwise (CW) and counter clockwise (CCW) spinning: the probability of CW rotation increase with the bacteria chemotaxis concentration (Che Y-P) [19]. When all the E.coli motors rotate CCW, flagella form coherent bundles and the cell runs linearly; when CW rotation is given by at least one motor, the flagella works independently and the cell rotates with little net displacements (tumbling). These two alternating rotation modes driven by Che-Y concentration suggested a two-state model of the single cell [5] with temporal variations of the energy barriers that couldn't account for the peaked structures in the power spectra as discussed above.

Optical noise arises from the flagellar motors spinnings activated by the phosphorated protein fluctuations of concentration: measurements detect these fluctuations of optical density in experiments lasting 5m. Reported spectra show prominent peaks centered at approximately 0.3 Hz and $1/f^g$ behaviors in the low frequency range. At growing chemotaxis concentration both the $1/f^g$ intensity and the γ values decreases whereas the peak appears more relevant. It turns out that the γ values are typically rather low, in the range 1-1.3. In these conditions, the intra avalanche correlation among elementary events is small and the correlation among large events becomes more relevant. In our frame, the elementary events are the individual switching events of flagellar motors which give detectable *1/f* noise when correlated in bundles whenever flagella rotate CCW. Finally, the motor spinning amplitude and the duration elapsing between two subsequent motor activations suggests to prefer distributions with $s<1$ which involves the relevance of short-time correlation. 3D stereoplots of E.coli swimming [22] show long running typically followed by significantly shorter displacements and vice-versa.

The choice around 0.5 gives the best fit of the experimental spectra, as shown in Fig.3. The agreement of the spectra calculated with our statistical model is quite satisfactory. As discussed above, the parameters $\rho\tau_0$, in the range *0.3 ¸ 0.5* in our fits, is related to the mean duration of the elementary events in bundles. Since Berg observed running and tumbling times of 1 and 0.1 s, respectively [22], our results suggest that the dynamics of the cell be averaged in the reported spectra.

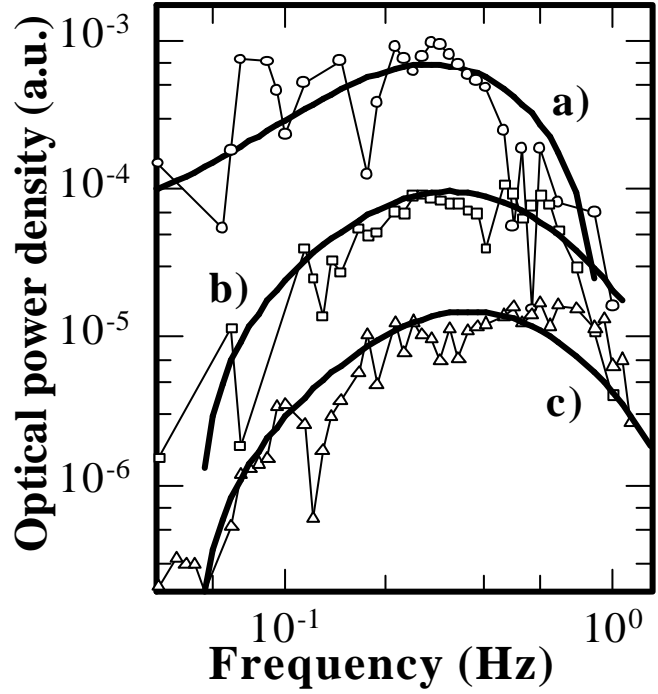

FIG.3 – Frequency dependence of the power density in the dynamics of the E.coli single cell. Experimental data are taken from E.Korobkova et al, ref. 4. From a) to c) spectra are collected at growing Che-Y concentration of the protein activating both CW rotations and optical activity. Solid curves are obtained from eqn.4: fit parameters in the various concentrations are in the ranges s = 0.5 ÷ 0.7 and $\rho\tau_0$ = 300 ÷ 500 ms. The peaks are obtained subtracting a background from the measured power spectra as in Fig.2. The three curves are staked for clarity in the figure.

*CONCLUSIONS*

Our analytical treatment explains quantitatively peaked structures observed in a variety of systems. The compelling suggestions of the recent literature that peaks are due to temporal correlation effects [3,5] are ultimately answered in the present work. The adopted time-amplitude correlation between large avalanches due to clustered sequences of elementary events results in a general analytical development suitable to be applied to power spectra calculations in a very wide frequency range from below 1Hz to above MHz for different natural systems.